\documentstyle[pra,aps,twocolumn,epsf]{revtex}
\begin{document}
\draft
\twocolumn[\hsize\textwidth\columnwidth\hsize\csname @twocolumnfalse\endcsname
\title{Multicritical Phenomena of Superconductivity and 
Antiferromagnetism in Organic Conductor $\kappa$-(BEDT-TTF)$_2$X}
\author{Shuichi Murakami \cite{author} and Naoto Nagaosa}
\address{Department of Applied Physics, University of Tokyo,
Bunkyo-ku, Tokyo 113-8656, Japan}
\date{\today}
\maketitle
\begin{abstract}
We study theoretically the multicritical phenomena of the 
superconductivity (SC) and antiferromagnetism (AF) in 
organic conductors  $\kappa$-BEDT salts.
The phase diagram and the experimental data on the NMR relaxation 
rate $1/T_1$ is analysed in terms of the renormalization 
group method.
The bicritical phenomenon observed experimentally 
     indicates the rotational symmetry, i.e., SO(5) symmetry,
     of the SC and the AF. 
The critical exponent $x$ for the divergence of $1/T_1$
     is well explained by $ x = \nu ( z - 1 - \eta)$
     with the dynamical exponent $z = 3/2$ for the AF region while 
     $z = \phi/\nu \sim 1.84 $ at the bicritical point.
These results strongly suggest that the origin of the SC is in common with
that of the AF and that its symmetry is d-wave.
\end{abstract}
\pacs{ 74.70.Kn, 74.20.De, 75.25.+z, 71.27.+a}

\vskip2pc]

\narrowtext
In strongly correlated electronic systems, it often happens that 
superconductivity and magnetically ordered states touch or are near to 
each other. Heavy electron superconductors \cite{hv1,hv2}, and
high-$T_{\rm c}$ superconductors \cite{htsc} are examples of this phenomenon,
where the Coulomb interactions are believed to be large compared with the 
bandwidth. In organic conductors, e.g., 
(TMTSF)$_2$X (X:anion) \cite{jerome} and $\kappa$-(BEDT-TTF)$_2$X 
\cite{williams,kanoda}, on the other hand, 
the magnitude of the Coulomb interaction is smaller
because the  molecular orbitals are extended.
However, the bandwidth $W$ is also small 
($\sim 1\ {\rm eV}$)
in these systems, and the electrons are half-filled when dimerization   
is considered.
Therefore, it is expected that the Coulomb interactions play important 
roles in these compounds. 
In particular, in  $\kappa$-(BEDT-TTF)$_2$X, recent experiments 
have revealed an interesting phase diagram including the 
bicritical phenomenon between the antiferromagnetism (AF) and the 
singlet superconductivity (SC)
in the plane of temperature and a parameter $p$ controlling the ratio of the 
Coulomb repulsion $U$ and the bandwidth $W$ \cite{williams,kanoda,kmnk,kmk}.
Above a characteristic temperature $T^*(U/W)$, 
the NMR relaxation rate, $T^{-1}_1$, increases as the 
temperature is lowered in a manner independent of $p$.
Below $T^*(U/W)$, on the other hand, $T_1^{-1}$ diverges towards the 
AF transition temperature, while it exhibits a spin-gap-like
behavior on the SC side. 

It is a crucial theoretical issue to treat AF and SC in a unified fashion 
in the physics of strongly correlated systems. One of the most interesting
proposals is based on the SO(5) symmetry \cite{zhang}, 
i.e., the rotational symmetry 
in the 5-dimensional order parameter space of AF and SC.
The critical phenomena are are useful for testing this idea because 
of their universality, and actually give the  first evidence for it in these
organic conductors. 

It is well known that the Ginzburg criterion 
$\frac{|T-T_{{\rm c}} |}{T_{{\rm c}}} <  (\frac{ a}{\xi})^2$
specifies the critical region in 3D where the mean field theory breaks down. 
Here, $a$ is the lattice constant and $\xi$ is the correlation length,
whose ratio is determined roughly as
${ a \over \xi } \sim { {\Delta } \over {W_{{\rm eff}}} }$,
where $\Delta$ is the gap introduced in the electronic spectrum by the AF 
and/or the SC, and $W_{{\rm eff}}$ is the reduced effective bandwidth 
by electron 
correlation. The mean field theory of the SDW for the 
AF and the BCS theory for the SC applies 
in the weak coupling regime, i.e., $\Delta \ll W_{{\rm eff}}$, 
and the critical 
phenomenon is dominated by the mean-field-like behavior in this case.
This occurs when the Coulomb interaction is weak ($ U \ll W $) for the AF
and/or the electron-phonon coupling is weak for the SC.   
In the strong coupling regime, on the other hand, 
we have a sufficiently large critical region and the critical
behavior and the phase diagram are determined by the fluctuations beyond the 
mean-field theory.

\begin{figure}[htbp]
\begin{center}
\vspace{0mm}
\hspace{0mm}
\epsfxsize=8.3cm
\epsfbox{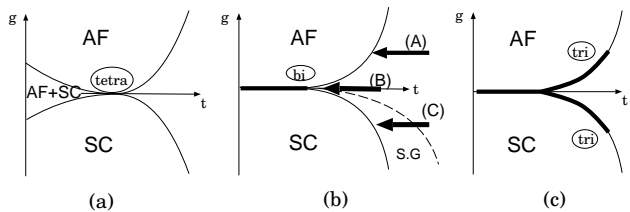}
\vspace{0mm}
\caption{Schematic phase diagrams of the model (\ref{model}) as a 
function of $t\sim(2r_{\|} +3r_{\bot})/5$ and $g\sim (r_{\|} -
r_{\bot})$. The thick and the thin lines represent first-order and 
second-order lines, respectively. 
``S.G.'' denotes a region with a spin-gap behavior.
(A)-(C) correspond to materials discussed later in the text.
(a) $uv>w^{2}$: tetracritical, 
(b) $uv=w^{2}$: bicritical, 
(c) $uv<w^{2}$: tricritical. }
\label{phase}
\end{center}
\end{figure}

In this paper, we study theoretically the critical phenomena of 
the AF and the SC in $\kappa$-(BEDT-TTF)$_2$X from the viewpoint that this
system is in the strong coupling regime and nontrivial 
critical phenomena are observed. 
The fluctuation of both the AF and the SC order parameters are treated 
in terms of the renormalization group (RG) method,
employing $\epsilon=4-d$ expansion. The critical phenomena occurs 
at finite temperature and hence is three dimensional. 
The anisotropy of the system is taken into account by proper rescaling 
and $\xi=\sqrt[3]{\xi_x \xi_y \xi_z}$, and the critical region is
discussed later. 
Classification of the scaling trajectories leads to
three types of multicritical phenomena (Fig.\ \ref{phase}): 
(a) tetracritical, (b) bicritical, and (c) tricritical phenomena.
The bicritical phenomenon (b) is unstable
towards tetra- and tricritical ones, and is realized only when there
is SO(5) symmetry.
Furthermore, the dynamic critical phenomena are studied, particularly for the 
NMR relaxation rate  $T_1^{-1}$, and its critical exponent $x$
is in reasonable agreement with the $\epsilon$-expansion result in both the 
AF region and at the bicritical point.
Thus, the analysis of multicritical phenomena gives quantitative
evidence of SO(5) symmetry to a good accuracy without adjustable parameters, 
and it strongly suggests 
that the SC is realized by the Coulomb interaction  as well as the AF and its 
pairing symmetry is d-wave.

 Let us consider a generic  Ginzburg-Landau model for a system with competing 
AF and SC order:
\begin{eqnarray}
&&H=\int d^{d}\mbox{\boldmath$r$} \left[
\frac{1}{2}r_{\|}|\vec{\sigma}|^{2}+
\frac{1}{2}|\vec{\nabla}\vec{\sigma}|^{2}+
\frac{1}{2}r_{\bot}|\vec{s}|^{2}+
\frac{1}{2}|\vec{\nabla}\vec{s}|^{2}\right. \nonumber \\
&&\ \ +
\left.
u|\vec{\sigma}|^{4}+
2w|\vec{\sigma}|^{2}|\vec{s}|^{2}
+v|\vec{s}|^{4}\right],
\label{model}
\end{eqnarray}
where $\vec{\sigma}$ and $\vec{s}$ are the order parameters
of the SC and the AF, respectively.
We normalize the order parameters so that the coefficients of 
the gradient terms are 1/2, and the other coefficients are
roughly given in the unit where $a=1, W=1$ by
$r_{\|} \cong { {( T - T_{{\rm c, SC}} )} 
\over { T_{{\rm c, SC}} \xi_{{\rm SC}}^2} }$, 
$r_{\bot} \cong { {( T - T_{{\rm c, AF}} )} \over { T_{{\rm c, AF}} 
\xi_{{\rm AF}}^2} } $,
$u \cong \xi_{{\rm SC}}^{-2} $, 
$v \cong \xi_{{\rm AF}}^{-2} $,
while $w$ represents the competition of the AF and the 
SC, and is expected to be
positive.
This model was proposed in refs.~\cite{bcmr,bl} in the context of
the SO(5) theory for high-$T_{{\rm c}}$ superconductors
\cite{zhang}, and was also used in \cite{kfs} for the heavy fermion compound 
Ce$_{x}$Cu$_2$Si$_2$ $(x\sim 1)$ to discuss the effect of disorder. 
This model has two types of mean field phase diagrams
\cite{lf}.
When the competition of the AF and the SC is not strong,
i.e., $ w^2 < u v $, there occurs the coexistence of the 
AF and SC, and the phase diagram looks like Fig.\ref{phase}(a). 
When $ w^2 > u v $,
the AF and the SC are separated by a first-order phase transition line and 
the phase diagram looks like Fig.\ \ref{phase}(b).
One might be tempted to interpret the phase diagram and
the bicritical phenomenon observed experimentally in terms of this mean
field picture. However, there are several experimental facts
which imply that the system is in the 
strong coupling regime at least for the AF:
(i) the large saturation value of the staggered magnetization 
$M_s \cong 0.4 \mu_B$ \cite{kanoda2}, which is nearly the value 
for the 2D Heisenberg AF,
(ii) fluctuation of the  staggered moments are observed as the broadening
of the NMR line shape well above $T_{{\rm c, AF}}$ \cite{kmnk} and
(iii) the spin-gap behavior is observed in $(T T_1)^{-1}$ well above 
     $T_{{\rm c, SC}}$.
Hence, the critical region near the bicritical point should be large
provided that $\xi_{{\rm AF}} \sim 1$.

Let us proceed to the analysis of the fluctuation 
in terms of the RG method. 
The RG recursion relations for $u,v,w$ up to order $\epsilon$ 
are written as \cite{nkf}
\begin{eqnarray}
&&\frac{du}{dl}= \epsilon u-\frac{(n_{\|}+8)u^{2}+n_{\bot}w^{2}}{2\pi^{2}}
\label{RG1}\\
&&\frac{dv}{dl}= \epsilon v-\frac{(n_{\bot}+8)v^{2}+n_{\|}w^{2}}{2\pi^{2}}
\label{RG2}\\
&&\frac{dw}{dl}= \epsilon w-\frac{(n_{\|}+2)u+(n_{\bot}+2)v+4w}{2\pi^{2}}w.
\label{RG3}
\end{eqnarray}
It should be noted that the recursion relations for $u,v,w$ do not involve 
$r_{\|},r_{\bot}$ up to this order.
There are six fixed points~\cite{nkf}, and
a stable fixed point depends on the number of components
$n_{\|}$ and $n_{\bot}$
of the vectors $\vec{\sigma}$ and $\vec{s}$.
If $n=n_{\|}+n_{\bot}$ is smaller than 
$n_{c}=4-2\epsilon+O(\epsilon^{2})$, a 
Heisenberg fixed point with $u=v=w\neq 0$ is the only stable one. 
This fixed point becomes unstable when $n$ exceeds $n_{c}$, 
and the so-called biconical fixed point with 
unequal values of $u$,$v$,$w$ becomes stable instead. 

\begin{figure}[htbp]
\begin{center}
\vspace{0mm}
\hspace{0mm}
\epsfxsize=5cm
\epsfbox{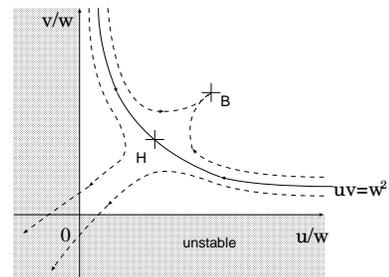}
\vspace{0mm}
\caption{Schematic diagram for the RG flow of the model (\ref{model}) in the 
$u/w$-$v/w$ plane. Broken lines represent the flow. Points ``B''
and ``H'' are the biconical and the Heisenberg fixed points, respectively.
The shaded region is the unstable region of the model.
}
\label{flow}
\end{center}
\end{figure}

In the present case with $n_{\|}=2,\ n_{\bot}=3$,
the only stable fixed point is the biconical one
$(u^{*}_{{\rm B}},v^{*}_{{\rm B}},
w^{*}_{{\rm B}})=2\pi^{2}\epsilon (0.0905,0.0847,0.0536)$.
However, not all points 
will flow to this fixed point through the RG (Fig.\ref{flow}).
There exists a curved surface, or a ``separatrix'', $F(u,v,w)=0$,
which divides the entire parameter space into two regimes:
one of convergence to the fixed point $(u^{*},v^{*},w^{*})$, 
and the other in which the RG flows run away into an unstable region.
Numerical analysis shows that this curved surface is well approximated 
by the surface $uv=w^{2}$ in the vicinity of the isotropic line
$u=v=w$; thus, we can roughly say $F(u,v,w)\sim uv-w^{2}$.

If $F(u,v,w)\sim 
uv-w^{2}>0$, the RG flow converges to the biconical fixed point 
$(u^{*}_{{\rm B}},v^{*}_{{\rm B}},w^{*}_{{\rm B}})$
and the phase diagram shows a tetracritical behavior (Fig.\ \ref{phase}(a)), 
as is predicted by the mean-field approximation~\cite{lf}.
From the RG recursion relations of $r_{\|}$ and $r_{\bot}$ with this 
biconical fixed point, we obtain $\nu_{{\rm B}}=0.5+0.132\epsilon$  
and $\alpha_{{\rm B}}=-0.0278\epsilon$.
The four second-order phase boundaries are 
written as $|g|\sim |t|^{\phi_{{\rm B}}}$ with 
the crossover exponent $\phi_{{\rm B}}=1+0.135\epsilon$,
where $t\sim \frac{2r_{\|}+3r_{\bot}}{5}$ and
$g\sim r_{\|}-r_{\bot}$ are scaling fields corresponding to 
the temperature and the anisotropy between the AF and
SC phases, respectively. We shift the definitions of $t$ and $g$ 
properly so that the tetracritical point corresponds to $t=g=0$.
The values of the exponents are different from the ones in \cite{bl}, 
which might be due to their approximation of
smallness of $\frac{n_{\bot}-n_{\|}}{n_{\bot}+n_{\|}}$.
A deviation of $(u,v,w)$ from the biconical fixed point is irrelevant
and affects only the size of the tetracritical region.

If $F(u,v,w)\sim uv-w^{2}<0$, on the other hand, although 
the mean-field approximation predicts a bicritical behavior~\cite{lf},
it is not the case. 
Through the RG flow, $u$ or $v$ decreases rapidly to 
become negative,
and the model (\ref{model}) becomes unstable. 
There are always higher-order terms which stabilize the system, 
though they are omitted in (\ref{model}).
Accordingly, the phase transition between the disordered and the ordered 
phases becomes first order (fluctuation-induced first-order 
transition \cite{1st}), but if the quadratic 
anisotropy $|g|$ is sufficiently large, the transition 
becomes second order again~\cite{dmf}. 
Thus, a first-order transition line between two ordered phases
branches at the triple point and extends until these two branches 
terminate at tricritical points (Fig.\ \ref{phase}(c)).
If $(u,v,w)$ is near the surface $uv=w^{2}$, the length of the 
first-order lines on the normal-SC or the normal-AF phase
boundaries will be small, since it takes a long time to flow into an 
unstable region.

Only in the case $uv=w^{2}$ does the RG analysis predict
a bicritical behavior (Fig.\ \ref{phase}(b)).
This behavior is governed by 
the Heisenberg fixed point 
$u_{{\rm H}}^{*}=v_{{\rm H}}^{*}=w_{{\rm H}}^{*}=
2\pi^{2}\epsilon/13$, which is stable only on the 
surface $uv=w^{2}$, and the corresponding exponents are 
$\nu_{{\rm H}}=0.5+0.135\epsilon,
\phi_{{\rm H}}=1+0.192\epsilon,
\alpha_{{\rm H}}=
-0.0385\epsilon$. 
The experimentally observed bicritical behavior,
combined with the above discussion, strongly suggests
$uv=w^{2}$ to a good accuracy,
which corresponds to the rotational symmetry in the 
five-dimensional order parameter space of 
$(\vec{\sigma}, \vec s)$.
It is interesting to note that only if $uv=w^{2}$ is the model 
(\ref{model}) smoothly related to 
the SO(5) NL$\sigma$ model in \cite{zhang,hkt}, 
in the limit $-r_{\|},-r_{\bot},u,v,w
\rightarrow\infty$ with their ratios fixed.

 Now let us consider the dynamic critical phenomena~\cite{hh}.
The NMR relaxation rate \cite{kmnk,kmk}
exhibits a divergence toward the transition in the AF side,
or a spin-gap behavior in the SC side, while these two 
cases show a similar behavior above $T^{*}$. This 
is reminiscent of the bicritical crossover behavior 
of the model (\ref{model}). 
The NMR linewidth is proportional to $(T-T_{c,{\rm AF}})^{-x}$ 
on the AF side of the normal phase with $x=\nu(z-1-\eta)$,
where $z$ is a dynamic critical exponent~\cite{hh}. 
When the system is in the vicinity of $T_{c, {\rm AF}}$, but not very 
near the bicritical point, its dynamic critical behavior is governed
by that of the AF isotropic Heisenberg model. Macroscopic 
varibles describing slow dynamics are the 
staggered magnetization $\vec{s}$ 
and the
conserved uniform magnetization $\vec{m}$, giving $z=d/2=3/2$.
Thus, the exponent is $x_{{\rm AF}}=0.315$ up to 
$O(\epsilon^{3})$ \cite{eps3}. 
On the other hand, when the bicritical fixed point 
governs the dynamic critical phenomena, the value of $z$ 
changes, and so does the exponent $x$.
In the bicritical region, the SC order parameters $\vec{\sigma}$
enter the set of macroscopic variables.
The free energy $F$ has a scaling form:
\begin{equation}
F(t,g,\vec{\sigma},\vec{s},\vec{m})
=t^{2-\alpha}\Phi\left(
\frac{g}{t^{\phi}},
\frac{\vec{\sigma}}{t^{\beta}},
\frac{\vec{s}}{t^{\beta}},
\frac{\vec{m}}{t^{\tilde{\beta}}}\right),
\end{equation}
where $\tilde{\beta}=2-\alpha-\phi_{{\rm H}}$.
$t$ is proportional to $T-T_{c,{\rm BP}}$, where $T_{c,{\rm BP}}$ denotes the 
temperature at the bicritical point.
Following the power-counting argument of the bicritical 
behavior in the spin-flop AF~\cite{hr}, we get 
$z=\phi_{{\rm H}}/\nu_{{\rm H}}\sim 1.84$, resulting in $x_{{\rm BP}}=0.584$, 
where we used $\phi_{{\rm H}}\sim 1.313$, $\nu_{{\rm H}}\sim 0.714$ 
up to $O(\epsilon^{2})$ \cite{eps3}. 
This value of $x_{\rm BP}$ at the bicritical point 
increases as a function of $n$:
$x_{\rm BP}=0.528, 0.558, 0.584, 0.607$ for $n=3,4,5,6$, respectively.
On the other hand, in approaching the SC phase it does not diverge 
but exhibits a spin-gap behavior below a characteristic temperature 
$t_{{\rm S.G.}}^{*}\sim|g|^{1/\phi_{{\rm H}}}$ 
because of the singlet formation. 
Throughout the entire critical region of the normal phase, $T_{1}^{-1}$ has a scaling 
form $T_{1}^{-1}=t^{-x_{{\rm BP}}} f(g/t^{\phi_{{\rm H}}})$.
\begin{figure}[htbp]
\begin{center}
\vspace{0mm}
\hspace{0mm}
\epsfxsize=5.5cm
\epsfbox{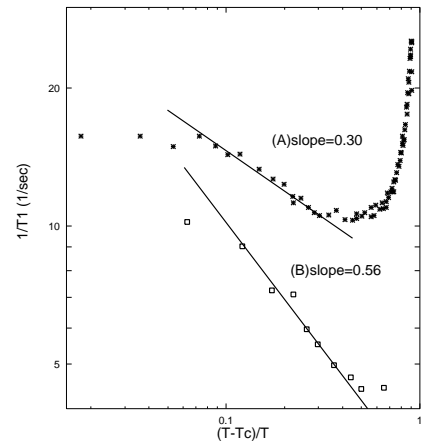}
\vspace{0mm}
\caption{
Log-log plot of $T_1^{-1}$ vs $(T-T_{{\rm c}})/T$
for (A) $\kappa$-(BEDT-TTF)$_{2}$Cu[N(CN)$_{2}$]Cl (solid squares),
and (B) deuterated $\kappa$-(BEDT-TTF)$_{2}$Cu[N(CN)$_{2}$]Br 
(open squares). (Data from \protect\cite{kmnk}.)}
\label{exp}
\end{center}
\end{figure}

Fig.\ \ref{exp} 
shows the log-log plot of $T_{1}^{-1}$ vs $\frac{T-T_{{\rm c}}}{T}$
for (A) $\kappa$-(BEDT-TTF)$_{2}$Cu[N(CN)$_{2}$]Cl (solid squares),
and (B) deuterated $\kappa$-(BEDT-TTF)$_{2}$Cu[N(CN)$_{2}$]Br (open squares).
The former is located in the AF region slightly distant from the bicritical point
while the latter is nearly at the bicritical point, as shown in Fig.\ref{phase}(b).
We selected $\frac{T-T_{c}}{T}$ instead of $\frac{T-T_{c}}{T_{c}}$ as
the abscissa, because the critical temperatures of these compounds 
are much lower than their classical values ($\sim J \sim 500\ {\rm K}$),
indicating that they are close to a quantum critical point.
This scaling by $\frac{T-T_{c}}{T}$ near the quantum critical point 
is obtained from an RG study of 
a quantum SO(5) NL$\sigma$ model with a quadratic anisotropy, and
a detailed discussion will be given in a subsequent paper \cite{mn}.
The proximity to the quantum bicritical point 
is also consistent with the strong correlation, 
because $T_{c}$ is suppressed 
by the quantum fluctuation driven by the Coulomb interaction.
The estimated slope, i.e., the critical exponent $x$, in the region
$1\ {\rm K} \lesssim T-T_{c, {\rm AF}} \lesssim 10\ {\rm K}$ 
gives $x_{{\rm AF}}\sim 0.30\pm 0.04$ for (A)
and $x_{{\rm BP}}\sim 0.56\pm 0.04$ for (B).
The values of $x$ are in reasonably good agreement with the above
theoretical values ($x_{{\rm AF}}= 0.314,\
x_{{\rm BP}}= 0.584$), which supports that 
the critical region is wide ($\sim
10\ {\rm K}$) and $\xi_{{\rm AF}}\sim 1$. 
Several points should be noted. 
(i) The fitting with $\frac{T-T_{c}}{T_{c}}$ 
underestimates the values of $x$ as
$x\sim 0.24$ for (A) and $x\sim 0.40$ for (B);
thus, it guarantees the validity of our fitting with 
$\frac{T-T_{c}}{T}$.
(ii) Note that the data in $T-T_{c}\lesssim 1\ {\rm K}$ 
seems to deviate from the above 
fitting. We discard these 
data in the fitting for the following 
reasons. First, $T_{c}$ cannot be determined more 
accurately than $\Delta T_{c}\sim 0.5\ {\rm K}$ 
due to experimental limitations. This
uncertainty largely affects the fitting within $T-T_{c}\lesssim 
1\ {\rm K}$. Second, in real systems, SO(5) symmetry is approximate,
and there is always a small 
deviation from $uv-w^{2}=0$, which 
alters the critical behavior very close to $T_{c}$.
Here, it is important to 
note that 
the increase of the deviation through the RG flow is slow;
in the linearization of (\ref{RG1})-(\ref{RG3}),
the exponent $\epsilon/13$ of the growth
is much smaller than the magnitude of the 
other two exponents $-8\epsilon/13, -\epsilon$.
Therefore, the bicritical behavior governed by the 
SO(5) fixed point can be observed for a relatively wide range of 
$|uv-w^{2}|$. For example, $|uv-w^{2}|\lesssim 0.1\cdot
{\rm max}(uv,w^{2})$ is sufficient to explain the observed bicritical 
phenomenon.

Let us consider whether it is possible to have such a large ($\sim
10\ {\rm K}$) critical region in the quasi-2D system. 
Following \cite{scalapino}, 
we can evaluate the critical region of the quasi-2D Heisenberg system 
as 
\begin{equation}
 \frac{|T-T_{{\rm c}}|}{T_{{\rm c}}}\lesssim \frac{kT_{{\rm c}}}{J_{\|}}\sim 
\frac{1}{\ln (J_{\|}/J_{\bot})},
\label{width}
\end{equation} 
where the in-plane 
exchange $J_{\|}$ is larger than 
the interplane exchange $J_{\bot}$. 
Thus there is only logarithmic dependence on $J_{\bot}/J_{\|}$, and
it does not significantly reduce the width of the critical region.
In $\kappa$-(BEDT-TTF)$_{2}$X in a metallic region, the anisotropy of 
the conductivity is $\sigma_{\|}/\sigma_{\bot}\sim 100$, implying 
$t_{\|}/t_{\bot}\sim \sqrt{100}$ and 
$J_{\|}/J_{\bot}\sim 100$. Thus, the width (\ref{width}) is not extremely 
small
($\sim
0.2-0.3$), and in the AF region ($T_{{\rm c}}\sim 30 {\rm K}$)
the observed
size of the critical region $5\ {\rm K}-10\ {\rm K}$ is reasonable.
In contrast, in quasi-1D systems, the width in (\ref{width}) is 
$J_{\bot}/J_{\|}$;
the quasi-one-dimensionality is effective in reducing 
the width of the critical region.

On the SC side, the experimental estimation of 
the coherence length $\xi_{{\rm SC}}$
is still controversial, e.g.,
for the SC compound $\kappa$-(BEDT-TTF)$_{2}$Cu(NCS)$_{2}$, 
the calculated in-plane coherence length 
$\xi_{{\rm SC}}$ ranges from $31\ \mbox{\AA}$
\cite{i} to $\sim 180\ \mbox{\AA}$ \cite{iy}.
However, 
the rotational symmetry between the AF and SC further suggests that the 
correlation length $\xi_{{\rm SC}}$ is also short ($\sim 1$).
This is because when $\xi_{{\rm SC}} \gg \xi_{{\rm AF}}$, 
the scaling trajectory in Fig.\ \ref{flow} 
starts from the initial point ($u \ll v$) 
far away from the Heisenberg fixed point, and the deviation 
from the separatrix is easily magnified as the length scale 
becomes larger.  
It is also inferred that the observed spin-gap behavior is due to 
the large fluctuation of the SC order parameter
in the SC compound $\kappa$-(BEDT-TTF)$_{2}$Cu(NCS)$_{2}$ ((C) in 
Fig.\ \ref{phase}(b)).

One might wonder that the easy-axis anisotropy 
reduces the number of the components of the AF order parameters
and stabilizes the bicritical phenomenon. However, 
the spin-orbit interaction is negligible in organic systems, and
the spin anisotropy energy due to dipole-dipole 
interactions is estimated by the AF resonance \cite{torr}
to be $\sim 10^{-4}\ {\rm K}$ for (TMTSF)$_2$X.
Assuming a similar value for $\kappa$-(BEDT-TTF)$_{2}$X, and 
that $\xi_{{\rm AF}} \sim 1$, this anisotropy is negligible in the
temperature region of interest.

Finally, the implications of these results are discussed below.
The rotational symmetry between AF and SC order parameters
suggest that the mechanism of the AF and SC is common \cite{km} and 
that 
the underlying microscopic quantum model has enhanced dynamical 
symmetry, i.e., SO(5) \cite{zhang}. Therefore, it is likely that the 
symmetry is d-wave. We believe that the (a) half-filling,
(b) intermediate Coulomb interaction, and (c) nearly two-dimensional
Fermi surface in $\kappa$-(BEDT-TTF)$_{2}$X makes the SO(5)
symmetric model a promising candidate.
When these conditions are largely violated, e.g., by
doping carriers and/or the application of an external magnetic field, 
the SO(5) symmetry is broken and the bicritical phenomenon  
turns into tetra- or tricritical behavior \cite{hkt}.

The authors are grateful to K.~Kanoda, M.~Kardar, and K.~Miyagawa
for fruitful discussions. 
This work is supported by a Grant-in-Aid for 
COE Research No.~08CE2003
from the Ministry of Education, Science, Culture and Sports of Japan.


\begin{thebibliography}{99}
\bibitem[*]{author} e-mail: murakami@appi.t.u-tokyo.ac.jp
\bibitem{hv1} G.~R.~Stewart, Rev. Mod. Phys. {\bf 56} 755 (1984).
\bibitem{hv2} P.~A.~Lee {\it et al.},
Comments in Solid State Physics {\bf 12} 99 (1986).
\bibitem{htsc} P.~W.~Anderson, Science {\bf 235} 1196 (1987).
\bibitem{jerome}
D.~J{\'e}rome, in {\it Organic Conductors}, ed. J.-P.~Farges,
(Marcel Dekker, New York, 1994) p.405.
\bibitem{williams}
J.~M.~Williams {\it et al.}, Science {\bf 252} 1510 (1991).
\bibitem{kanoda}
K.~Kanoda, Hyperfine Interactions {\bf 104} 235 (1997).
\bibitem{kmnk}
A.~Kawamoto, K.~Miyagawa, Y.~Nakazawa and K.~Kanoda,
Phys. Rev. {\bf B52} 15522 (1995); unpublished.
\bibitem{kmk}
A.~Kawamoto, K.~Miyagawa and K.~Kanoda, Phys. Rev. {\bf B55}
14146 (1997).
\bibitem{zhang}S.-C.~Zhang, Science {\bf 275} 1089 (1997).
\bibitem{bcmr}
C.~P.~Burgess {\it et al.}, Phys. Rev. {\bf B57} 8549 (1998).
\bibitem{bl}
C.~P.~Burgess and C.~A.~L{\"u}tken, Phys. Rev {\bf B57} 8642 (1998).
\bibitem{kfs}
H.~Kohno, H.~Fukuyama and M.~Sigrist, J. Phys. Soc. Jpn.
{\bf 68} 1500 (1999).
\bibitem{lf}
K.~-S.~Liu and M.~E.~Fisher, J. Low Temp. Phys. {\bf 10} 655 (1972).
\bibitem{kanoda2}
K.~Kanoda, Physica {\bf C282-287} 299 (1997).
\bibitem{nkf}
D.~R.~Nelson, J.~M.~Kosterlitz, and M.~E.~Fisher, 
Phys. Rev. Lett. {\bf 33} 813 (1974);
J.~M.~Kosterlitz, D.~R.~Nelson, and M.~E.~Fisher, 
Phys. Rev. {\bf B13} 412 (1974).
\bibitem{1st}
T.~Natterman and S.~Trimper, J. Phys. {\bf A8} 2000 (1975);
S.~A.~Brazovskii and I.~E.~Dzyaloshinskii, JETP Lett. {\bf 21} 164 (1975);
J. Rudnick, Phys. Rev. {\bf B18} 1406 (1978).
\bibitem{dmf}
E.~Domany, D.~Mukamel and M.~E.~Fisher, Phys. Rev. {\bf B15} 5432 (1977)
\bibitem{hkt}
X.~Hu, T.~Koyama and M.~Tachiki, Phys. Rev. Lett {\bf 82} 2568 (1999);
X.~Hu, cond-mat/9906237.
\bibitem{hh}
P.~C.~Hohenberg and B.~I.~Halperin, Rev. Mod. Phys. {\bf 49} 435 (1977).
\bibitem{eps3}
M.~E.~Fisher, Rev. Mod. Phys. {\bf 46} 597 (1974);
K.~G.~Wilson and J.~Kogut, Phys. Rep. {\bf C12} 75 (1974).
\bibitem{hr}
D.~L.~Huber and R.~Raghavan, Phys. Rev. {\bf B14} 4068 (1976).
\bibitem{mn}
S.~Murakami and N.~Nagaosa, in preparation.
\bibitem{scalapino}
D.~J.~Scalapino, Y.~Imry and P.~Pincus, Phys. Rev. {\bf B11} 2042 (1975).
\bibitem{i}
H.~Ito {\it et al.}, J. Phys. Soc. Jpn. {\bf 60} 3230 (1991).
\bibitem{iy}
K.~Oshima {\it et al.}, J. Phys. Soc. Jpn. {\bf 57} 730 (1988);
K.~Murata {\it et al.}, Synth. Metals {\bf 27} A341 (1988).
\bibitem{torr}
J.~B.~Torrance, H.~J.~Pedersen, and K.~Bechgaard,
Phys. Rev. Lett. {\bf 49} 881 (1982).
\bibitem{km}
H.~Kondo and T.~Moriya, J. Phys. Soc. Jpn. {\bf 68} 3170 (1999).
\end{thebibliography}
\end{document}